\documentclass[preprint,aps,nofootinbib]{revtex4}
\usepackage{}
\usepackage{graphicx}
\usepackage{amsmath}
\usepackage{amsfonts}
\usepackage{amssymb}
\usepackage{color}%
\usepackage{dcolumn}
\usepackage{indentfirst}
\providecommand{\U}[1]{\protect\rule{.1in}{.1in}}

\definecolor{lightgray}{rgb}{.7,.7,.7}

\definecolor{red}{rgb}{1,0,0}

\definecolor{blue}{rgb}{0,0,1}

\definecolor{purple}{rgb}{0.6,0.1,0.7}

\newcommand{\f}{\begin{equation}}
\newcommand{\ff}{\end{equation}}
\newcommand{\fa}{\begin{eqnarray}}
\newcommand{\ffa}{\end{eqnarray}}

\begin{document}
\title{Note on quantum entanglement and quantum geometry}
\author{Yi Ling $^{1,2}$}
\email{lingy@ihep.ac.cn}
\author{Yikang Xiao $^{1,2}$}
\email{ykxiao@ihep.ac.cn}
\author{Meng-He Wu $^{1,2}$}
\email{mhwu@ihep.ac.cn} \affiliation{$^1$Institute of High Energy
Physics, Chinese Academy of Sciences, Beijing 100049, China\\ $^2$
School of Physics, University of Chinese Academy of Sciences,
Beijing 100049, China}

\begin{abstract}
In this note we present preliminary study on the relation
between the quantum entanglement of boundary states and the
quantum geometry in the bulk in the framework of spin networks. We
conjecture that the emergence of space with non-zero volume
reflects the non-perfectness of the $SU(2)$-invariant tensors.
Specifically, we consider four-valent vertex with identical spins
in spin networks. It turns out that when $j=1/2$ and $j=1$, the
maximally entangled $SU(2)$-invariant  tensors on the
boundary  correspond to the eigenstates of the volume square
operator in the bulk, which indicates that the quantum geometry of
tetrahedron has a definite orientation.
\end{abstract}
\maketitle

\section{Introduction}
Recently more and more evidences have been accumulated to support
the conjecture that the geometric connection of spacetime is just
the emergent phenomenon of the quantum entanglement style of
matter, which has been becoming an exciting arena for the
interaction of quantum information, quantum gravity and condensed
matter
physics\cite{Maldacena:2001kr,Ryu:2006bv,VanRaamsdonk:2010pw,Vidal:2007hda,Swingle:2009bg,Swingle:2012wq}.
In particular, in AdS/CFT approach, the  relation between
the minimal surface in the bulk and the entanglement entropy for
boundary states has been quantitatively described by the
Ryu-Takayanagi formula, which is recently understood from the
quantum error correction (QEC) scenario as
well\cite{Harlow:2016vwg}. In this approach the perfect tensor
network plays a key role in mimicking the function of QEC
for hyperbolic space\cite{Pastawski:2015qua}. Here the notion of
perfectness means that the entanglement entropy could saturate the
maximal value which is given by the local degrees of freedom on
the boundary, for {\it any} bipartition of particles in which the
smaller part contains particles no more than half of the total
particles. Among all the kinds of tensor networks, perfect tensor
network exhibits the strongest ability of QEC, in the sense that
information can always be recovered by pushing it from the bulk
towards the boundary in all directions. Unfortunately, the tensor
network built with perfect tensors always exhibits a flat
entanglement spectrum, which is not consistent with the
holographic nature of AdS space, which is characterized by the
non-flat entanglement spectrum. Recent work in
\cite{Ling:2018vza,Ling:2018ajv} indicates that in order to have a
non-flat entanglement spectrum one has to sacrifice the ability of
tensors for QEC, which implies that the tensors in network should
not be perfect if a non-flat entanglement spectrum is expected to
achieve.

Based on the above progress, it is quite intriguing to investigate
the relation between quantum entanglement of boundary states and
the geometric structure of the bulk in a non-perturbative way,
since the holographic nature of gravity has widely been
accepted as the fundamental principle for the theory of quantum
gravity. Preliminary explorations on the entanglement entropy of
boundary states in the framework of spin networks have appeared in
literature\cite{Orus:2014poa,Han:2016xmb,Li:2016eyr,Li:2017vvh,Chirco:2017vhs,Chirco:2017xjb,Livine:2017fgq,Baytas:2018wjd,Ling:2018yaj}.
 In this framework, gauge invariant quantum states
play a key role in describing the quantum geometry of polyhedrons.
In particular, intertwiners as $SU(2)$-invariant tensors are basic
ingredients for the construction of spin network states, which is
proposed to describe the quantum geometry of space time as well as
the quantum states of gravitational field in four dimensions. To
investigate the QEC in AdS space which is supposed to be described
by quantum geometry at the  microscopic level, it is quite
interesting to discuss the perfectness of the boundary states in
the framework of spin networks. Recently, it is shown in
\cite{Li:2016eyr} that bivalent and trivalent tensors can be
both $SU(2)$-invariant and perfect which are uniquely given by
the singlet state or $3j$ symbols. However,  for n-valent
tensors when $n$ is four or more than four, it is not possible to
construct a $SU(2)$-invariant  tensor that is perfect at the
same time (unless the spin $j$ is infinitely large, which is
called asymptotically perfect tensors in
\cite{Li:2016eyr,Li:2017vvh}). This is a very interesting result
because it is well known in spin network literature that the
volume operator has non-zero eigenvalues only when acting on vertices
with four or more
edges\cite{Rovelli:1994ge,Rovelli:1995ac,Loll:1995wt}. That is to
say, when a $SU(2)$-invariant tensor is perfect, the corresponding
volume of space must be vanishing. Based on this fact, we
conjecture that {\it the emergence of the space with non-zero
volume is the reflection of the non-perfectness of
$SU(2)$-invariant tensors}.

In this note we intend to find more features of $SU(2)$-invariant
tensors and then disclose the relation between the quantum
entanglement of boundary states and the quantum geometry in the
bulk. In particular, we propose a quantity to measure the
non-perfectness of a single $SU(2)$-invariant tensor. For a
boundary state, we define the sum of the entanglement entropy over
all the possible bipartition as $S_{tot}$. Then the
non-perfectness of any tensor can be evaluated by the difference
between $S_{tot}$ and that of a perfect tensor $S_{p}$, which
is uniquely determined by the number of degrees of freedom on the
boundary. We may denote it as $\Delta S$.  The corresponding state
with the maximal value of $S_{tot}$ is called as the maximally
entangled state. If $\delta=\Delta S/S_{p}$ is tiny, then this
maximally entangled state may be called as {\it nearly} perfect
tensor\footnote{A similar notion for random invariant tensors
rather than a single tensor is introduced in \cite{Li:2017vvh}.}.
In this note we intend to find these maximally entangled states on
the boundary and consider their relations with the quantum states
in the bulk for the simple spin network which only contains a
single vertex with four dangling edges, describing a quantum
tetrahedron geometrically. Correspondingly, the boundary state is a
4-valent tensor state.

Our main result is that when $j=1/2$ and $j=1$, the maximally
entangled $SU(2)$-invariant tensors on the boundary correspond to
the eigenstates of the square of the volume operator in the bulk,
which indicates that the geometry of quantum tetrahedron has a
definite orientation. This paper is organized as follows. In next
section we present the setup for four-valent $SU(2)$-invariant
tensors and give the boundary states with the maximal entanglement
entropy for $j=1/2$ and $j=1$, while the detailed derivation of
these states is presented in Appendix. Then the  relation between
these states and the quantum states of the tetrahedron in the bulk
is given in section III. Our numerical results on the relations
between the entanglement entropy and the expectation of the volume
for general states is given in section IV.  Section V is the
conclusions and outlooks.

\section{The boundary states with the maximal entanglement entropy}

\begin{figure} [h]
  \center{
  \includegraphics[scale=0.5]{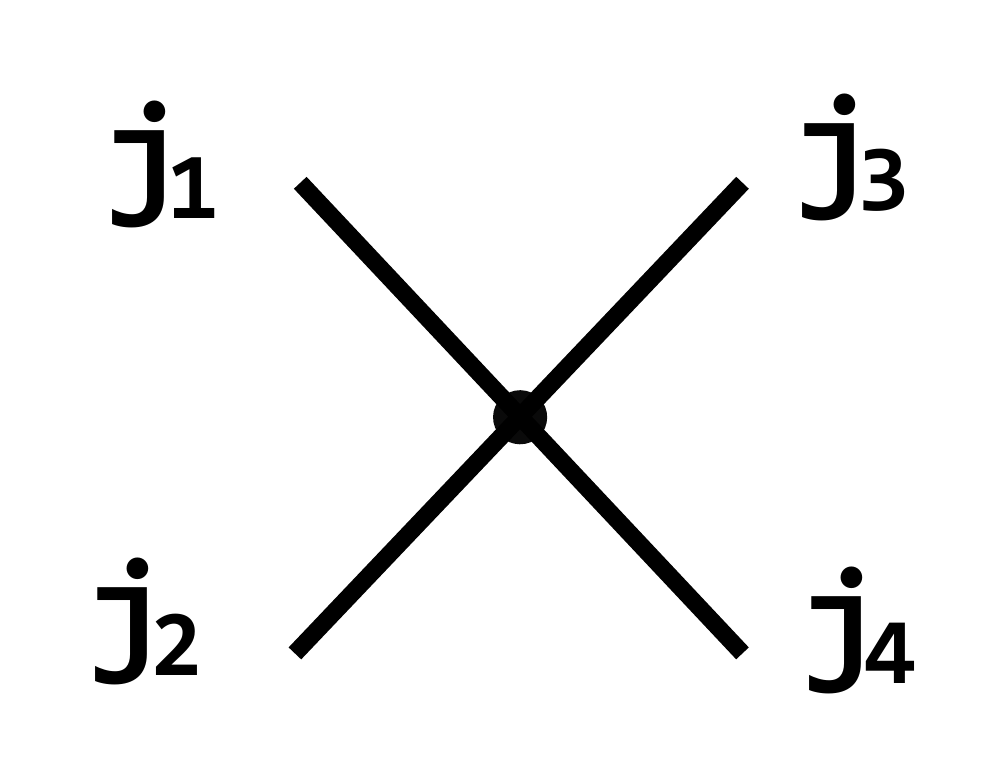}\ \hspace{0.05cm}
  \caption{\label{fig1} The sketch of a 4-valent $SU(2)$-invariant tensor with single vertex.}}
\end{figure}

The setup is given as follows. We consider a 4-valent tensor associated with a
single vertex, which can be diagrammatically sketched as
Fig.\ref{fig1}.

To be perfect or almost perfect for any bipartition, we only
consider the case that all the external legs are identically
labelled by spin $j$, namely $j_i=j(i=1,2,3,4)$, then a 4-valent
tensor can be written as
\begin{equation}
\left |\psi_4  \right \rangle =\sum_{m_1 m_2 m_3 m_4}\psi_{m_1 m_2 m_3 m_4 } \left|m_1 m_2 m_3
m_4 \right \rangle  ,
\end{equation}
where $m_i=-j,-j+1,...,j-1,j.$ To be
$SU(2)$-invariant, we know $\left |\psi_4 \right \rangle$ must be a singlet satisfying
$\sum_{i}j_i \left|\psi_4 \right \rangle =0$. As a result, we find the tensor states
must have the following form
\begin{eqnarray}
\psi_{m_1 m_2 m_3 m_4 } &=& \sum_{J}\alpha(J) \left \langle   m_1 m_2 m_3 m_4 | J \right \rangle  \nonumber \\
&=& \sum_{J}\frac{\alpha(J)}{\sqrt{2J+1}}\sum_ {M_1 M_2}C_{m_1 m_2
JM_1}^{(j_1 j_2)} C_{m_3 m_4 JM_2}^{(j_3 j_4)} D^{(J)M_1 M_2 }
\label{jbzk} ,
\end{eqnarray}
where $C_{m_1 m_2 \ J \ M}^{(j_1  j_2)} = \left \langle j_1 \ m_1
;j_2 \ m_2 | J \ M\right \rangle$ is the standard Clebsch-Gordan
$(CG)$ coefficient and $D^{(J)M_1 M_2}$ is a $(2J+1)\times (2J+1)$
matrix with $D^{(J)M_1 M_2}=(-1)^{J-M_1}\delta^{M_1,-M_2}$.
Moreover, the possible value of $J$ is determined by the coupling
rules of two spins, here running as $0,...,2j-1,2j$. $\alpha(J)$
are free complex numbers which are specified by the intertwiner.

Next we consider the entanglement entropy with bipartition. Since
the entanglement entropy for  the $(1,3)$ bipartition is trivial
which is identically $\ln(2j+1)$, we only need to consider the
bipartition with equal legs in each part. If two external
legs of 4-valent tensor are combined and labelled by a single
index, then tensors can be treated as matrices. For instance, if
$j_1$ and $j_2$ are paired, then the reduced density matrix is
given by
\begin{equation}
\rho_{12} = \sum_{m_1 m_2 m_1'm_2'm_3 m_4}\psi_{m_1 m_2 m_3 m_4 } \psi_{m_1' m_2' m_3 m_4}^{*} \left | m_1 m_2 \right \rangle \left \langle  m_1' m_2'\right |.
\end{equation}

Since the tensor is a pure state, one has $\rho_{12}=\rho_{34}$.
For four-valent tensor, there are three ways to pair the external
legs. Thus, the corresponding entanglement entropy for
bipartition can be calculated as
\begin{equation}
S_{12}=-tr\rho_{12} \ln\rho_{12},\ \ \ \ \ S_{13}=-tr\rho_{13} \ln\rho_{13},\ \ \ \ S_{14}=-tr\rho_{14} \ln\rho_{14}.
\end{equation}

In \cite{Li:2016eyr} it is proved that 4-valent $SU(2)$-invariant
tensors can not be perfect, in the sense that it is not possible
to construct a state $|\psi_4\rangle$ such that the entanglement
entropy saturates the bound $S_{12}=S_{13}=S_{14}=2\ln (2j+1)$. In
another word, if the entanglement entropy $S_{12}=S_{13}=2\ln
(2j+1)$, then the entanglement entropy $S_{14}$ must be less than
$2\ln (2j+1)$. Based on this fact, then it is quite natural to ask
what kind of $SU(2)$-invariant tensors could be nearly perfect, in
the sense that it is maximally entangled among all the
$SU(2)$-invariant tensors. Next we intend to provide an answer to
this issue by figuring out the $SU(2)$-invariant tensor with the
maximal entanglement entropy for some specific spin $j$. For
4-valent tensors, Such a nearly perfect tensor is defined as the
state with the maximal value for the sum of the entanglement
entropy, namely $S_{tot}=S_{12}+S_{13}+S_{14}$.

Firstly, we consider the simplest case with $j=1/2$. Our goal is
to find $\alpha(0)$ and $\alpha(1)$ such that $S_{tot}$ takes the maximal value. In appendix, we analytically show
that when $\alpha(0)=\pm i\alpha(1)$, the entanglement entropy
takes the  maximal value, which is given by
\begin{eqnarray}
S^m_{tot}=Max(S_{12} + S_{13} + S_{14}) = 3 \ln(2\sqrt{3}).
\end{eqnarray}
For a perfect tensor, this value is expected to be $S_{p}=3\ln4$. Thus we find the ``deficit'' of the entanglement entropy is $\Delta S=S_{p}-S^m_{tot}=3\ln(2/\sqrt{3})\simeq 0.43$, and
$\delta\simeq 0.104$. We remark that it is interesting to notice that the second order
of Renyi entropy $S^{(2)}_{ik} = - \ln tr(\rho _{ik}^{2})$ takes
the maximal value as well
\begin{eqnarray}
Max(S^{(2)}_{12} + S^{(2)}_{13} + S^{(2)}_{14}) = 3\ln3.
\end{eqnarray}
We notice that the entanglement spectrum is not flat for these
maximally entangled states indeed, unlike the perfect tensors.
Ignoring the global phase factor, the corresponding states are
\begin{equation}
\psi_{m_1 m_2 m_3 m_4 }=\frac{1}{\sqrt{2}}  \left \langle    m_1
m_2 m_3 m_4 |J=0 \right \rangle  \pm\frac{i}{\sqrt{2}} \left
\langle   m_1 m_2 m_3 m_4 |J=1 \right \rangle  . \label{j12}
\end{equation}

Next we consider the case of $J = 1$. In parallel, we find two
states having the maximal entanglement entropy. The corresponding
intertwiner $\sum_{J}\alpha(J) \left |J  \right \rangle $ is given by
\begin{equation}
\frac{\sqrt{2}}{3} \left |J=0  \right \rangle
\pm\frac{i}{\sqrt{2}} \left | J=1  \right \rangle
-\frac{1}{3}\sqrt{\frac{5}{2}} \left |J=2  \right \rangle .
\end{equation}

The corresponding entanglement entropy is
\begin{eqnarray}
S^m_{tot}=Max(S_{12} + S_{13} + S_{14})= \frac{5}{3}\ln2+\frac{9}{2}\ln3.
\end{eqnarray}
Thus the deficit of the entanglement entropy is $\Delta S=S_{p}-S^m_{tot}=6\ln3-(\frac{5}{3}\ln2+\frac{9}{2}\ln3)\simeq
0.49$, and $\delta\simeq 0.075$.

\section{The eigenstates of the volume operator on spin networks}

In this section we focus on the geometric interpretation of
invariant tensors with the maximal entanglement entropy. A
classical polyhedron in $R^3$ can be parameterized by the oriented
face area vectors subject to the closure condition. Quantum
mechanically, loop quantum gravity provides a well-known strategy
to quantize the polyhedrons based on spin network states, which are
$SU(2)$-invariant. The quantum volume operator can be defined by
quantizing the classical expression of the volume for a
three-dimensional region R, which is expressed in terms of Ashtekar variables as
\begin{equation}
V=\int d^3x\sqrt{g}=\int
d^3x\sqrt{\frac{1}{6}\epsilon^{ijk}\epsilon_{abc}E_i^a E_j^b
E_k^c},
\end{equation}
where $a,b,c$ are spatial indices, while $i,j,k$ are internal indices. In literature there exists two different strategies to construct
the volume operator and discuss its action on spin networks.
Traditionally, one is called the internal algorithm proposed by
Rovelli and
Smolin\cite{Rovelli:1994ge,Rovelli:1995ac,DePietri:1996tvo}, and
the other one is the external algorithm proposed by Ashtekar and
Lewandowski\cite{Ashtekar:1997fb,Thiemann:1996au}. In this paper only 4-valent
vertex is taken into account and these two versions are
equivalent\cite{Giesel:2005bk,Giesel:2005bm,Yang:2015wka}.

Before discussing the volume spectrum of 4-valent vertex, we
firstly elaborate our conjecture, arguing that the emergence of
the space with non-zero volume is the reflection of the
non-perfectness of $SU(2)$-invariant tensors. It is well known
that when the volume operator acts on any tri-valent vertex in
spin networks, the eigenvalue has uniformly to be zero. That is to
say, if a network only contains tri-valent vertices, the total
volume of the space corresponding to this state must be zero as
well. In this situation, perfect $SU(2)$-invariant tensors can in
principle be constructed based on this spin network. Specifically,
as investigated in \cite{Li:2016eyr}, for a tri-valent vertex
associated with three edges labelled by spins $j_1,j_2,j_3$, then
the $SU(2)$-invariant perfect tensor state is uniquely given by
Wigner's $3j$ symbols
\begin{equation}
\left |\psi_3  \right \rangle =\sum_{m_1 m_2
m_3}\begin{pmatrix}j_1&j_2&j_3\\m_1&m_2&m_3\end{pmatrix}\left|m_1m_2
m_3 \right \rangle.
\end{equation}
Then the total  $SU(2)$-invariant perfect tensor associated with a
network can be constructed by considering the products of these
 individual perfect tensors associated with each vertex. However, if one intends to construct a
space with non-zero volume with spin networks, four-valent or more
valent vertex must be included. Following the results in
\cite{Li:2016eyr}, then the $SU(2)$-invariant tensor associated
with this vertex has to be non-perfect. If some components or even
a single vertex of the network become non-perfect, we know that
the total $SU(2)$-invariant tensor based on the whole network can
not be perfect, either. Therefore, the emergence of non-zero
volume must accompany the non-perfectness of $SU(2)$-invariant
tensors in this scenario. Obviously, all the cases considered in
the remainder of this paper are subject to the conjecture that we
have proposed, because for all states with non-zero volume the
corresponding tensors are not perfect, indeed. More importantly,
next we will push this qualitative conjecture forward by
quantitatively demonstrating the relation between the value of the
volume and the value of maximal entanglement entropy for 4-valent
vertex.

For a 4-valent vertex, the action of the volume operator can be
described as $\hat{V}=\sqrt{l_p^6 |\hat{W}|}$, where
$l_p=\sqrt{8\pi G}$ is the planck length and for convenience we
set it as unit in the remaining part of this note. Here we
also remark that there is an overall coefficient which is
undetermined in the volume operator, but one can choose
appropriate coefficient such that the action of the volume
operator has a semiclassical limit correctly, as discussed in
\cite{Giesel:2005bk, Giesel:2005bm}. Nevertheless, this overall
coefficient does not affect our analysis in present paper on the
relation between entanglement and geometry. The operator
$\hat{W}$ is
\begin{equation}
\hat{W}=(1/8)\epsilon^{ijk} (\hat{J}_i^{(1)} \hat{J}_j^{(2)}
\hat{J}_k^{(3)}-\hat{J}_i^{(1)} \hat{J}_j^{(2)}
\hat{J}_k^{(4)}+\hat{J}_i^{(1)} \hat{J}_j^{(3)}
\hat{J}_k^{(4)}-\hat{J}_i^{(2)} \hat{J}_j^{(3)} \hat{J}_k^{(4)} ),
\end{equation}
where $\hat{J}_i^{(p)}$ represents the action of an angular
momentum operator $\hat{J}_i$ on the p-th edge associated with
the vertex.  $\hat{J}_i^{(p)}$ is the quantization of the smeared
triad $\frac{1}{2}\int_{M_{2}^{(p)}}\epsilon _{abc}
E_{i}^{a}dx^{b}dx^{c}$, and $M_{2}^{(p)}$ is 2-dimensional open
manifold which only intersects with the p-th edge once. The
detailed analysis about the action of the operator $\hat{W}$ on
intertwiners can be found in \cite{DePietri:1996tvo}, with  the
power of $6j$ and $9j$ symbols. Here for the case of 4-valent
vertex, one can find that matrix elements of $\hat{W}$ in
intertwiner space, namely $ \left \langle J' \right |\hat{W} \left
|J \right \rangle$, satisfy the rule $\Delta J=\pm 1$, i.e. $\left
\langle J'\right |\hat{W}\left |J \right \rangle \neq 0$ if and
only if $|J'-J|=1$. In addition, by virtue of the Hermitian of the
operator, one has $Re \left \langle J+1 \right |\hat{W}\left |J
\right \rangle =Re \left \langle J \right |\hat{W} \left |J+1
\right \rangle =0$, and $Im \left \langle J+1 \right |\hat{W}
\left |J \right \rangle=-Im \left \langle J \right \rangle
|\hat{W} \left |J+1 \right \rangle$.

Now, for 4-valent vertex with $j= 1/2$, we have
\begin{equation}
\hat{W} = \left(
  \begin{array}{ccc}
0 & -\frac{\sqrt{3}}{8}i \\
 \frac{\sqrt{3}}{8}i  &  0
  \end{array}
\right).
\end{equation}

The eigenvalues of $\hat{W}$ are $\pm \sqrt{3}/8$, corresponding
to the eigenstates $\left|\pm \right \rangle=\frac{1}{\sqrt{2}}
\left |J=0 \right \rangle \pm \frac{i}{\sqrt{2}}\left | J=1 \right
\rangle$, respectively. In literature, these two eigenstates are
understood as the quantum states of the  tetrahedron with definite
orientation, where $\left |+ \right \rangle$ corresponds to the
right-handed orientation, while $\left |- \right \rangle$
corresponds to the left-handed orientation.  Surprisingly, we find
that these eigenstates  are nothing but giving rise to the
4-valent $SU(2)$-invariant  states with maximal entanglement
entropy on the boundary. It is worthwhile to understand the
geometric interpretation of this correspondence. First of all, in
intertwiner space, no matter what values $\alpha(0)$ and
$\alpha(1)$ are, the spin network states are always eigenstates of
the volume operator $\hat{V}$, with the eigenvalue of
$(\sqrt{3}/8)^{\frac{1}{2}}$, but the orientation of the
tetrahedron is usually mixed. Only the eigenstates of the operator
$\hat{W}$ have a definite orientation. Therefore, in this simplest
case with $j=1/2$, we find that the boundary states with the
maximal entanglement entropy correspond to the quantum states of
the tetrahedron  with definite orientation.

Moreover, it is interesting to understand the emergence of non-zero eigenvalue of the volume from the viewpoint of quantum information. In \cite{Li:2016eyr}, it is shown that the tri-valent $SU(2)$-invariant tensors can be perfect, implying that the quantum information could be recovered by QEC with full fidelity. On the other hand, it is known that the action of the volume operator  on any tri-valent vertex gives rise to the zero eigenvalue of the volume.  Once the volume of the polyhedron is non-zero, like the operator acting on four-valent vertex, then the $SU(2)$-invariant tensor can not be perfect any more,  implying that the quantum information must sacrifice or lose its fidelity when teleporting through the vertex for some certain partitions. Or conversely, one can say that in order to guarantee  the polyhedron, as the basic bricks of space,  has non-zero volume, then as the channel of QEC, the $SU(2)$-invariant tensor can not be perfect. In a word, the space with non-zero volume emerges as the deficit of the entanglement entropy, or the loss of the fidelity of QEC. This is the  key observation in this note.

Next we consider the case of $j=1$, then the intertwiner space is
spanned with $\left |J=0 \right \rangle,\left |J=1 \right
\rangle$, and $\left |J=2 \right \rangle$. The matrix $\left \langle
J' \right |\hat{W}\left |J \right \rangle$ reads as
\begin{equation}
\hat{W} = \left(
  \begin{array}{ccc}
0 & -\frac{i}{\sqrt{3}} & 0 \\
 \frac{i}{\sqrt{3}}  &  0 & -\frac{i}{2}\sqrt{\frac{5}{3}} \\
 0 & \frac{i}{2}\sqrt{\frac{5}{3}} & 0
  \end{array}
\right).
\end{equation}

It turns out that $\hat{W}$ has two non-zero eigenvalues $\pm \sqrt{3}/2$, corresponding to eigenstates
\begin{equation}
\left |\lambda_{\pm}\right \rangle=\frac{\sqrt{2}}{3}\left
|J=0\right \rangle\pm \frac{i}{\sqrt{2}}\left |J=1\right
\rangle-\frac{1}{3}\sqrt{\frac{5}{2}}\left |J=2\right \rangle,
\end{equation}
as well as an eigenvalue 0, corresponding to the eigenstate
\begin{equation}
\left |\lambda_{0}\right \rangle=\frac{\sqrt{5}}{3}\left |J=0\right \rangle+\frac{1}{2}\left |J=2\right \rangle.
\end{equation}
Remarkably, we find that the boundary states with the maximal
entanglement entropy obtained in previous section correspond to
the quantum state of tetrahedron with definite orientation in the
bulk.  It is worthwhile to point out that for general intertwiner
parameters, the states are not the eigenstates of the volume
operator any more, but the expectation value can be evaluated. In
next section we intend to investigate the relation between the
entanglement entropy of boundary states and the volume or orientation of the
tetrahedron by numerical analysis.

\section{Numerical results}
In this section we present the relation between the entanglement
entropy of boundary states and the volume and orientation of
tetrahedron in the bulk by randomly selecting the parameters in
intertwiner space.

For $j=1/2$, a general state $\left | \psi _{4}  \right \rangle$ in intertwiner space can be expanded as
\begin{eqnarray}
\left | \psi _{4}  \right \rangle &=& \alpha  \left | +   \right \rangle + \beta \left | - \right \rangle ,
\end{eqnarray}
where $\alpha,\beta$ are two  complex numbers.

\begin{figure} [h]
  \center{
  \includegraphics[scale=0.27]{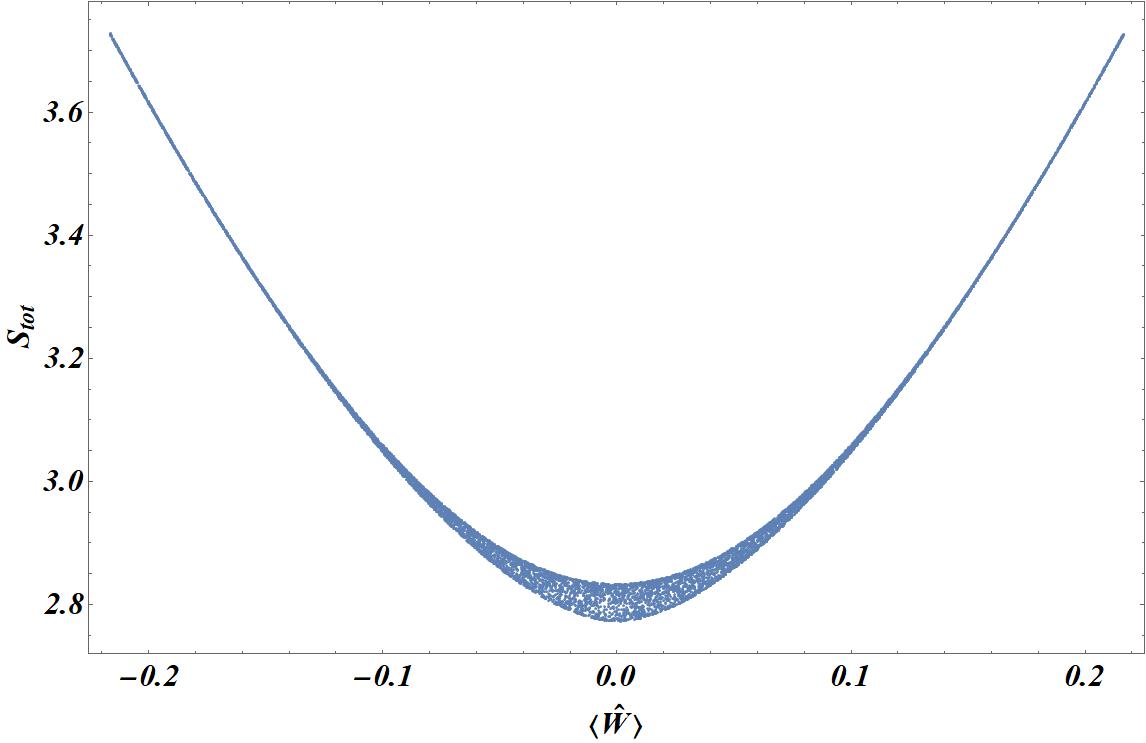}\ \hspace{0.05cm}
  \caption{\label{SW1} The relation between $S_{tot}=S_{12} + S_{13} + S_{14}$ and $ \langle \hat{W} \rangle$.}}
\end{figure}

We know that the volume operator $\hat{V}=\sqrt{|\hat{W} |}$, and
$\hat{W}\left | + \right \rangle= \frac{\sqrt{3}}{8}\left | +
\right \rangle $, $\hat{W}\left | - \right \rangle=
-\frac{\sqrt{3}}{8}\left | - \right \rangle $, then
\begin{eqnarray}
\langle \hat{W} \rangle &=& \frac{ \langle  \psi _{4}|  \hat{W}|  \psi _{4} \rangle}{ \langle  \psi _{4}  |  \psi _{4}  \rangle}= \frac{\sqrt{3}}{8}\frac{\left | \alpha  \right |^2 - \left | \beta  \right |^2}{\left | \alpha  \right |^2 + \left | \beta  \right |^2} , \nonumber \\
 \langle \hat{V}  \rangle &=& \frac{ \langle \psi _{4}|  \hat{V}| \psi _{4} \rangle}{ \langle \psi _{4}  | \psi _{4}  \rangle}= \left(\frac{\sqrt{3}}{8}\right)^{\frac{1}{2}}.
\end{eqnarray}
In Fig.\ref{SW1}, we show the relation between $S_{tot}$ and
$\langle \hat {V}  \rangle$ by randomly selecting complex numbers
$\alpha$ and $\beta$. From this figure, we justify that $S_{tot}$
does have the maximal value $3\ln(2\sqrt{3})$ when $\langle
\hat{W}  \rangle =\pm \sqrt{3}/8 $, which correspond to
$(\alpha=0, \beta=1)$ and $(\alpha=1,\beta=0)$, respectively. We
also notice that  $S_{tot}$  takes the minimal value at the
position with $ \langle \hat{W} \rangle=0$, which implies that the
geometry is the coherence of two oriented tetrahedron states
with equal probability, namely $|\alpha|=|\beta|$. In is also
interesting to notice that among all random states, a large
proportion of states distributes in the vicinity of $\langle
\hat{W} \rangle=0$ with lower entanglement entropy.  With the
increase of $|\langle \hat{W} \rangle|$, the proportion of states
becomes small but the entanglement entropy becomes larger. The
maximal value of entanglement entropy measures the ability of the
vertex as the channel of QEC, which is not perfect and consistent
with our conjecture. Furthermore, as the channel of QEC, it sounds
reasonable that the quantum tetrahedron with a definite
orientation has the maximal entanglement entropy, because its
deficit of entanglement entropy is the smallest such that it
possesses the best fidelity for quantum information
teleportation.

Similarly, we consider this relation for $j=1$. The general state
$\left | \psi _{4}  \right \rangle$ reads as
\begin{eqnarray}
\left | \psi _{4}  \right \rangle &=& \alpha _+\left | \lambda_+   \right \rangle + \alpha _0\left | \lambda_0   \right \rangle  + \alpha _-\left | \lambda_-   \right \rangle  ,
\end{eqnarray}
where $\alpha_\pm,\alpha_0$ are complex numbers. Then we have
\begin{eqnarray}
\langle \hat{W} \rangle &=& \frac{ \langle  \psi _{4}|  \hat{W}|  \psi _{4} \rangle}{ \langle  \psi _{4}  |  \psi _{4}  \rangle}= \frac{\sqrt{3}}{2} \frac{\left | \alpha _+ \right |^2 - \left | \alpha _- \right |^2}{\left | \alpha _+ \right |^2 + \left | \alpha _0 \right |^2 + \left | \alpha _- \right |^2} , \nonumber \\
 \langle \hat{V}  \rangle &=& \frac{ \langle \psi _{4}|  \hat{V}| \psi _{4}  \rangle}{\left \langle \psi _{4}  | \psi _{4}  \right \rangle}= \left(\frac{\sqrt{3}}{2}\right)^{\frac{1}{2}} \frac{\left | \alpha _+ \right |^2 + \left | \alpha _- \right |^2}{\left | \alpha _+ \right |^2 + \left | \alpha _0 \right |^2 + \left | \alpha _- \right |^2}.
\end{eqnarray}

\begin{figure} [h]
  \center{
  \includegraphics[scale=0.45]{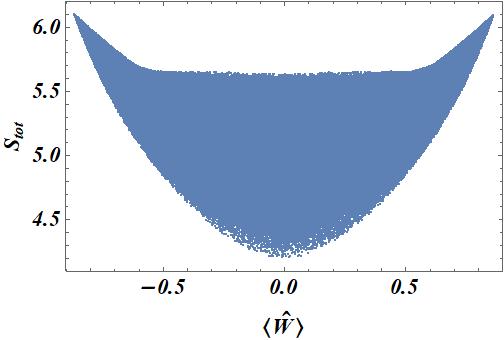}\ \hspace{0.05cm}
  \includegraphics[scale=0.45]{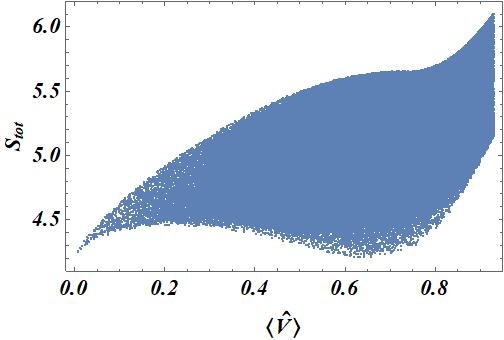}\ \hspace{0.05cm}
  \caption{\label{SWv2} Left: The relation between $S_{tot}$ and $ \langle \hat{W}  \rangle$. Right:  The relation between $S_{tot}$ and $ \langle \hat{V} \rangle$.}}
\end{figure}

In Fig.\ref{SWv2}, we show the relation between $S_{tot}$ and
$\langle \hat{W}  \rangle$, $\langle \hat{V} \rangle$ with random
numbers in intertwiner space. Again, our statement that $S_{tot}$
takes the maximal value for the eigenstates of  $\langle \hat{W}
\rangle$  is justified. In this case the relation between
$S_{tot}$ and  $ \langle \hat{V}  \rangle$ becomes complicated.
But it is true that the maximal value of $S_{tot}$ appears when
the expectation value of the volume takes the largest value. In
addition, when the expectation value of the volume is zero,
$S_{tot}$ takes the minimum.  In Fig.\ref{SWv2}, all the
corresponding $SU(2)$-invariant tensors are not prefect, even for
the states with zero volume. This result further indicates that
the quantum information has to lose its fidelity when teleporting
through a four-valent $SU(2)$-invariant tensor.

In the end of this section we point out that the minimal
deficit of the entanglement entropy $\Delta S \simeq 0.43$
for $j=1/2$, with the eigenvalue of $\langle \hat{V}
\rangle=(\sqrt{3}/8)^{\frac{1}{2}}$, is smaller than  $\Delta S
\simeq
0.49$
for $j=1$, with the eigenvalue of $\langle \hat{V} \rangle=
2(\sqrt{3}/8)^{\frac{1}{2}}$. This indicates that to build a
quantum space with larger volume, the minimal deficit of
entanglement entropy has to become larger as well. Intuitively, it
implies that more information has to be stored in the space to
form a space with larger volume. This observation could extend our
previous conjecture to a more quantitative version: The space with
non-zero volume must be built with non-perfect tensors.
Furthermore, the larger the volume is, the more deficit of
entanglement entropy for
non-perfect tensors one has to pay.

\section{Conclusions and outlooks}
In this note we have investigated the relations between the
entanglement entropy of the boundary states and the geometric
property of the quantum tetrahedron in the bulk for a single
4-valent vertex in the framework of spin networks. Qualitatively,
we have conjectured that the emergence of the space with non-zero
volume is the reflection of the non-perfectness of
$SU(2)$-invariant tensors. Based on this conjecture, we might
ascribe the increase or decrease of the space volume to the change
of the entanglement among particles on the boundary. Inspired by
this conjecture, it is quite interesting to explore the dynamics
of space from the side of the evolution of entanglement at the
Planck scale, for instance, at the beginning of the universe or
the cosmological inflation scenario where the quantum effect of
geometry becomes severe. Quantitatively, we have found the
relation between the maximally entangled states and the
eigenstates of the volume square operator. Interestingly enough,
we have found that for $j=1/2$ and $j=1$, the boundary
$SU(2)$-invariant states with the maximal entanglement entropy
correspond to the eigenstates of the operator $\hat{W}$, which
implies that the quantum tetrahedron has a definite orientation.
It is intriguing to ask whether this correspondence also holds for
other spins $j$. Our preliminary attempt indicates that for $j\geq
3/2$ there does not exist such simple relations between the states
with the maximal entanglement entropy and the eigenstates of the
operator $\hat{W}$. Their complicated relations deserve for
further investigation.

Although $j=1/2$ and $j=1$ are just the specific cases for  a
four-valent  states, this simple but elegant  correspondence  has
significant implications  for  understanding the deep relations
between the entanglement and the microscopic structure of the
spacetime, particularly in a non-perturbative manner.  As a
microscopic scenario of quantum spacetime,  representations  of
$j=1/2$ and $j=1$ are just like the ground state and the first
excited state of the system, which should be  dominantly occupied
among all the possible distributions.  This conjecture plays a key
role in the original work on the microscopic interpretation on the
entropy of black holes in terms of spin network
states\cite{Ashtekar:1997yu}.

The most desirable work next is to investigate the relations
between quantum entanglement and quantum geometry in the framework
of spin networks with more general setup.  We expect to compute
the entanglement entropy of a general boundary state, and explore
its dependence on the orientation of the quantum polyhedrons in
the bulk geometry. In this case, the main difficulty one faces is
the involvement of holonomy along edges. Since the volume operator
 non-trivially acts only on intertwiner space at vertices, it is
quite straightforward to discuss the geometric property of
polyhedrons, but in general the entanglement of boundary states
depends on the holonomy along edges, as previously studied in
\cite{Ling:2018yaj}. Our investigation on this topic is under
progress.

\section*{Acknowledgments}

We are very grateful to Yuxuan Liu and Zhuoyu Xian for helpful
discussions and suggestions. This work is supported by the Natural
Science Foundation of China under Grant No. 11575195 and 11875053.
Y.L. also acknowledges the support from Jiangxi young scientists
(JingGang Star) program and 555 talent project of Jiangxi
Province.

\section*{Appendix}
In this appendix we present the background on the spin network states with boundary, and then derive the $SU(2)$-invariant state with the maximal entanglement entropy for $j=1/2$. Given the connection $A$ and edge $e$, holonomy is defined as $h_e
=Pexp \int_e A$.  In irreducible representation of $SU(2)$, its matrix element is
\begin{eqnarray}
R_{mn}^j (h_e )=\left \langle j,n \right |h_e \left |j,m \right \rangle.
\end{eqnarray}

A closed spin network state can be represented by
$\left |\Gamma,{j_e },{I_v} \right \rangle$, where $\Gamma$ is the graph composed of edges $e$, labelled by the representation $j_e$, and vertices $v$, labelled by the intertwiner $I_v$. The relationship between the spin network representation and the
connection representation is given by
\begin{eqnarray}
 \left \langle h_{e_1},...,h_{e_N}  |\Gamma, \{j_e \},\{I_v\} \right \rangle = \sum_{\{m_e,n_e\}}\prod_{e\in E(\Gamma)}R_{m_en_e}^{j_e}(h_e)\prod_{v\in V(\Gamma)}(I_v)_{\{m_e,n_e\}}^{\{j_e\}}.
\end{eqnarray}

A spin network state with boundary can be represented by
$\left |\Gamma, {j_e,j_l },{I_v},{n_l} \right \rangle$, with  $e$ for inner edges, $l$ for
dangling edges where the magnetic quantum number $n_l$ is specified. The relationship between spin network
representation and the connection representation is
\begin{eqnarray}
 \left \langle \{h_{e}\},\{h_{l}\}|\Gamma, \{j_e,j_l \},\{I_v\},\{n_l\} \right \rangle&=&\sum_{m_e,n_e,m_l}\prod_{e\in E(\Gamma)}R_{m_en_e}^{(j_e)}(h_e)  \nonumber \\
 &\times& \prod_{v\in V(\Gamma)}(I_v)_{\{m_e n_e,m_l\}}^{\{j_e,j_l\}}\prod_l R_{m_l n_l}^{j_l}(h_l).
\end{eqnarray}

Once $\Gamma$  and ${j_e}$ are specified, a spin
network state with boundary can also be written as
\begin{eqnarray}
\left |\Gamma ,\{j_e,j_l \},\{I_v \},\{n_l \} \right \rangle=\bigotimes _v  \left | I_v^{\{j_e,j_l\}} \right \rangle \bigotimes _l \left \langle j_l,n_l \right |.
\end{eqnarray}

Thus $\left \langle \{h_e\},\{h_l\} \right |\bigotimes  \left |{I_v} \right \rangle$ can be mapped to the right vector,
$\left|\psi(\{h_e\},\{h_l\},\{I_v\})\right \rangle$, which is given as
\begin{eqnarray}
\left|\psi(\{h_e\},\{h_l\},\{I_v\})\right \rangle
&=&\sum_{n_l}\psi_{\{n_l\}}(\{h_e\},\{h_l\},\{I_v\})\prod_{l} \left |\{j_l\},\{n_l\}\right \rangle,
\end{eqnarray}
where
\begin{eqnarray}
\psi_{n_l}(\{h_e\},\{h_l\},\{I_v\})&=&\sum_{n_l'}\psi_{\{n_l'\}}(\{h_e\},\{h_l=I\},\{I_v\})R_{n_l'n_l}^{j_l}(h_l),\\
\psi_{n_l}(\{h_e\},\{h_l=I\},\{I_v\})&=&\sum_{m_e,n_e,m_l,n_l}\prod_{e\in E(\Gamma)}R_{m_en_e}^{j_e}(h_e)\prod_{v\in V(\Gamma)}(I_v)_{\{m_e, n_e, m_l\}}^{\{j_e,j_l\}}\prod_l\delta_{m_l n_l},
\end{eqnarray}
with $I$ being the identity matrix. Usually, due to the presence of the boundary, the gauge invariance is broken. In this paper, we consider  a simple network which only contains a single vertex associated with
four dangling edges, so there is no $h_e$ and only one $I_v$ involved. The state is
\begin{eqnarray}
&&\psi_{n_1 n_2 n_3 n_4}(h_1,h_2,h_3,h_4,I_v) \nonumber \\ &&=\sum_{n_1'n_2'n_3'n_4'}\psi_{n_1'n_2'n_3'n_4'}(I,I,I,I,I_v )R_{n_1'n_1}^{j_1}(h_1)
 R_{n_2'n_2}^{j_2}(h_2)R_{n_3'n_3}^{j_3}(h_3)R_{n_4'n_4}^{j_4}(h_4)\nonumber\\
&&=\sum_{n_1'n_2'n_3'n_4'}(I_v )_{n_1'n_2'n_3'n_4'}^{j_1 j_2 j_3 j_4
}R_{n_1'n_1}^{j_1}(h_1)
R_{n_2'n_2}^{j_2}(h_2)R_{n_3'n_3}^{j_3}(h_3)R_{n_4'n_4}^{j_4}(h_4),
\end{eqnarray}
where $j_i=j_{l_i},n_i=n_{l_i} (i=1,...,4)$ and $\psi_{n_1'n_2'n_3'n_4'}(I,I,I,I,I_v )$ is a singlet and $SU(2)$ invariant.

Now we consider the entanglement entropy for such a 4-valent state with spin $j$. Without loss of generality, we consider the reduced density matrix by tracing the first and second index, leading to
\begin{eqnarray}
&&(\rho_{34})_{n_3 n_4 n_3'n_4'} (h_1,h_2 ,h_3,h_4,I_v )\nonumber\\
&&=\sum_{n_1 n_2}\sum _{n_1''n_2''n_3''n_4''}\psi_{n_1'' n_2''n_3''n_4''}(I,I,I,I,I_v )R_{n_1''n_1}^{j_1} (h_1 ) R_{n_2''n_2}^{j_2} (h_2 ) R_{n_3''n_3}^{j_3} (h_3 ) R_{n_4''n_4}^{j_4} (h_4 )\nonumber\\
&&\times\sum _{n_1'''n_2'''n_3'''n_4'''}\psi_{n_1 '''n_2'''n_3'''n_4'''}^* (I,I,I,I,I_v )  R_{n_1'''n_1}^{j_1*} (h_1 ) R_{n_2'''n_2}^{j_2*}
(h_2 ) R_{n_3'''n_3'}^{j_3*} (h_3 ) R_{n_4'''n_4'}^{j_4*} (h_4 ).
\end{eqnarray}

We remark that for a single vertex, the entanglement entropy  does  not
depend on $(h_1, h_2, h_3, h_4)$ because  all the reduced density matrices are related by similarity transformations.  Thus one can simply set them be
identity matrix $I$, leading to the reduced density matrix for a $SU(2)$-invariant tensor. Applying Eq.(\ref{jbzk}), one has
\begin{eqnarray}
(\rho_{34} )_{n_3 n_4 n_3'n_4'} (I_v )\propto\sum _{J=0}^{2j}\frac{|\alpha(J)|^2}{2J+1} \sum _MC_{n_3 n_4 JM_1}^{(jj)}  C_{n_3'n_4'JM_1}^{(jj)}.
\end{eqnarray}
Due to the unitarity of $CG$ coefficients, one can show that
\begin{eqnarray}
\rho_{34}\sim \frac{1}{\sum _{J=0}^{2j}|\alpha(J)|^2 } diag \left[(|\alpha(0)|^2)^{\oplus 1},(\frac{1}{3}|\alpha(1)|^2)^{\oplus 3},...,(\frac{1}{2j+1}|\alpha(j)|^2)^{\oplus(2j+1)}\right].
\end{eqnarray}
Then we derive the entanglement entropy as
\begin{eqnarray}
S_{34}=(\sum _{J=0}^{2j}|\alpha(J) |^2 )^{-1} (-\sum _{J=0}^{2j}|\alpha(J) |^2 \ln  \frac{|\alpha(J) |^2}{2J+1})+\ln\sum _{J=0}^{2j}|\alpha(J) |^2 .
\end{eqnarray}
Similarly, $ \left |\psi_4 \right \rangle$ can also be expanded based on other basis in intertwiner space as
\begin{eqnarray}
\psi_{m_1 m_2 m_3 m_4 }=\sum _J\frac{\beta(J)}{\sqrt{2J+1}} \sum _{M_1 M_2}C_{m_1 m_3 JM_1}^{(j_1 j_3)} C_{m_4 m_2 JM_2}^{(j_4 j_2)} D^{(J)M_1 M_2 }, \label{eq:ab}\\
\psi_{m_1 m_2 m_3 m_4 }=\sum _J\frac{\gamma (J)}{\sqrt{2J+1}} \sum
_{M_1 M_2}C_{m_1 m_4 JM_1}^{(j_1 j_4)} C_{m_2 m_3 JM_2}^{(j_2
j_3)} D^{(J)M_1 M_2 },
\end{eqnarray}
which is very convenient for us to calculate the entanglement entropy for other bipartitions. Specifically, we have
\begin{eqnarray}
S_{24}=(\sum _{J=0}^{2j}|\beta(J) |^2 )^{-1} (-\sum _{J=0}^{2j}|\beta(J) |^2 \ln\frac  {|\beta(J) |^2}{2J+1})+\ln\sum _{J=0}^{2j}|\beta(J) |^2, \nonumber\\
S_{23}=(\sum _{J=0}^{2j}|\gamma (J) |^2 )^{-1} (-\sum _{J=0}^{2j}|\gamma (J) |^2 \ln\frac  {|\gamma (J) |^2}{2J+1})+\ln\sum _{J=0}^{2j}|\gamma (J) |^2.
\end{eqnarray}
Next we will determine the values of parameters $\alpha(J), \beta(J), \gamma(J)$ such that the sum of the entanglement entropy  will take the maximal value among all the possible states. First of all, since $\alpha(J), \beta(J), \gamma(J)$ are parameters in different representations of the same state, they must be related to one another, we intend to derive their relations at first. One can easily find that $ \left \langle \psi_4 |\psi_4 \right \rangle=\sum
_{J=0}^{2j}|\alpha(J) |^2 =\sum _{J=0}^{2j}|\beta(J) |^2=\sum
_{J=0}^{2j}|\gamma (J) |^2=:U$, obviously $U>0$. Therefore, the sum of entanglement entropy reads as
\begin{equation}
S_{tot}=U^{-1} (-\sum _{J=0}^{2j}|\alpha(J) |^2 \ln \frac {|\alpha(J)
|^2}{2J+1}-\sum _{J=0}^{2j}|\beta(J) |^2 \ln \frac {|\beta(J)
|^2}{2J+1}-\sum _{J=0}^{2j}|\gamma (J) |^2 \ln \frac {|\gamma (J)
|^2}{2J+1})+3\ln U.
\end{equation}
From Eq. (\ref{jbzk}) and Eq. (\ref{eq:ab}), one has
\begin{eqnarray}
&&\sum _{J'}\frac{\alpha(J')}{\sqrt {2J'+1}} \sum _{M_1'M_2'}C_{m_1 m_2 J'M_1'}^{(j_1 j_2)} C_{m_3 m_4 J'M_2'}^{(j_3 j_4)} D^{(J')M_1 'M_2'}  \nonumber\\
&=&\sum _J\frac{\beta(J))}{\sqrt {2J+1}} \sum _{M_1 M_2}C_{m_1 m_3 JM_1}^{(j_1 j_3)} C_{m_4 m_2 JM_2}^{(j_4 j_2)} D^{(J)M_1 M_2 }.
\end{eqnarray}
From this equality, one can derive the following equation
 \begin{eqnarray}
&&\sum_{m_1 m_2 m_3 m_4}\sum_{J'}\frac {\alpha(J') }{2J+1}\sum_{M_{1'}M_{2'}}C_{m_1m_2J'M_1'}^{(j_1j_2)}C_{m_3m_4J'M_2'}^{(j_3j_4)}D^{(J')M_1'M_2'}C_{m_1m_3J''M_1''}^{(j_1j_3)*}C_{m_4m_2J''M_2''}^{(j_4j_2)*}\nonumber\\
&=&\sum_{m_1 m_2 m_3 m_4}\sum_{J}\frac  {\beta(J)
}{2J+1}\sum_{M_{1}M_{2}}C_{m_1m_3JM_1}^{(j_1j_3)}C_{m_4m_2JM_2}^{(j_4j_2)}D^{(J)M_1M_2}C_{m_1m_3J''M_1''}^{(j_1j_3)*}C_{m_4m_2J''M_2''}^{(j_4j_2)*}\nonumber\\
&=& \sum_{J} \frac{\beta (J)}{2J+1} \sum_{ M_1,M_2} \delta _{J}^{J''}\delta _{M_1}^{M_1''} \delta _{J}^{J''}\delta _{M_2}^{M_2''}D^{(J)M_1M_2}   \nonumber\\
&=&\frac {\beta(J'') }{2J''+1}D^{(J'')M_1''M_2''}.
\end{eqnarray}
Therefore, the term,  $C_{m_1 m_2 JM_1'}^{(j_1 j_2)} C_{m_3 m_4 JM_2'}^{(j_3
j_4)}D^{(J')M_1'M_2'}C_{m_1 m_3 J''M_1}^{(j_1 j_3)*} C_{m_4 m_2
J''M_2}^{(j_4 j_2)*}$, has $SU(2)$-invariance for $M_1,M_2$.  So let
\begin{equation}
C_{m_1 m_2 J'M_1'}^{(j_1 j_2)} C_{m_3 m_4 J'M_2'}^{(j_3
j_4)}D^{(J')M_1'M_2'}  C_{m_1 m_3 JM_1}^{(j_1 j_3)*}C_{m_4 m_2
JM_2}^{(j_4 j_2)*}=N(j,J',J)D^{(J)M_1M_2}.
\end{equation}
Then, we find the intertwiner parameters in different representations are related by
\begin{equation}
\beta(J)=\sqrt { 2J+1}\sum_{J'}N(j,J',J)\frac{\alpha(J)}{\sqrt {2J'+1}}.
\end{equation}
For the same reason, one has
\begin{equation}
\gamma (J)=\sqrt{2J+1}\sum_{J'}N'(j,J',J)\frac{\alpha(J)}{\sqrt{2J'+1}},
\end{equation}
where $N'(j,J',J)$ satisfies
\begin{equation}
C_{m_1 m_2 J'M_1'}^{(j_1 j_2)}C_{m_3 m_4 JM_2'}^{(j_3 j_4)}D^{(J')M_1' M_2' } C_{m_1 m_4 JM_1}^{(j_1 j_4)*} C_{m_2 m_3 JM_2}^{(j_2 j_3)*}=N'(j,J',J)D^{(J)M_1 M_2 }.
\end{equation}
We point out that $N(j,J',J),N'(j,J',J)$ can be explicitly
calculated by $6j$ symbols.
\begin{eqnarray}
N(j,J',J)=(2J'+1) (-1)^{J'}\begin{pmatrix}j&j&J\\j&j&J'\end{pmatrix},\nonumber\\
N'(j,J',J)=(2J'+1) (-1)^J \begin{pmatrix}j&j&J\\j&j&J'\end{pmatrix} .
\end{eqnarray}
Above equations give the general relations for parameters $\alpha(J), \beta(J), \gamma(J)$. Now we focus on the simple cases with specific spin $j$.
When $j= 1/2$, then $J=0,\ 1$. The non-trivial $6j$ symbols are
\begin{eqnarray}
\begin{pmatrix}\frac{1}{2}&\frac{1}{2}&0\\\frac{1}{2}&\frac{1}{2}&0\end{pmatrix}=-\frac{1}{2}, \ \begin{pmatrix}\frac{1}{2}&\frac{1}{2}&0\\\frac{1}{2}&\frac{1}{2}&1\end{pmatrix}=\frac{1}{2}, \
\begin{pmatrix}\frac{1}{2}&\frac{1}{2}&1\\\frac{1}{2}&\frac{1}{2}&0\end{pmatrix}=\frac{1}{2}, \ \begin{pmatrix}\frac{1}{2}&\frac{1}{2}&1\\\frac{1}{2}&\frac{1}{2}&1\end{pmatrix}=-\frac{1}{6}, \
\end{eqnarray}
which give rise to the following relations for $\alpha(J), \beta(J), \gamma(J)$
\begin{eqnarray}
\beta(0)=-\frac{1}{2} \alpha (0)-\frac{\sqrt {3}}{2}\alpha (1), \
\beta(1)=\frac{\sqrt {3}}{2}\alpha (0)-\frac{1}{2} \alpha (1), \nonumber\\
\gamma(0)=-\frac{1}{2} \alpha (0)+\frac{\sqrt {3}}{2}\alpha (1), \
\gamma(1)=-\frac{\sqrt {3}}{2}\alpha (0)-\frac{1}{2} \alpha (1) .\label{relation}
\end{eqnarray}
As a result, we find the sum of the entanglement entropy is
\begin{eqnarray}
S_{tot}&=&U^{-1} (-|\alpha (0) |^2 \ln |\alpha (0) |^2-|\beta(0) |^2 \ln |\beta(0) |^2-|\gamma(0) |^2 \ln |\gamma(0) |^2\nonumber\\ &-&|\alpha (1) |^2 \ln |\alpha (1) |^2/3-|\beta(1) |^2 \ln |\beta(1) |^2/3-|\gamma(1) |^2\ln |\gamma(1) |^2/3)+3\ln U.
\end{eqnarray}

Furthermore, from Eq.(\ref{relation}), one can derive that
\begin{eqnarray}
|\alpha(1)|^2+|\beta(1)|^2+|\gamma(1)|^2=|\alpha (0)|^2+|\beta(0)|^2+|\gamma(0)|^2=\frac{3}{2} (|\alpha (0) |^2+|\alpha (1) |^2 )=\frac{3}{2} U.
\end{eqnarray}

By virtue of  the inequality $\ln x\leq x-1 (x>0)$, one  can show that
\begin{eqnarray}
-|\alpha(0)|^2\ln(|\alpha(0)|^2) &\leq& \frac{U}{2}-\ln(\frac{U}{2})|\alpha(0)|^2,\nonumber\\
|\alpha (1) |^2 \ln |\alpha (1) |^2/3 &\leq& U/2-|\alpha
(1) |^2-\ln(U/6) |\alpha (1) |^2.
\end{eqnarray}
Similar inequality can be derived for $\beta$ and $\gamma$. Where the equal sign of the inequality holds if and only
if $|\alpha (0) |^2=|\beta(0) |^2=|\gamma(0) |^2=|\alpha (1)
|^2=|\beta(1) |^2=|\gamma(1) |^2=U/2$, which leads to
\begin{equation}
\alpha (1)/\alpha (0) =\pm i, \ \  with \  \  \alpha (0)\neq0.
\end{equation}

So ignoring the phase factor, there are two maximally entangled states. They are
\begin{equation}
\frac{1}{\sqrt{2}} \left |J=0 \right \rangle \pm\frac{i}{\sqrt{2}} \left |J=1\right \rangle.
\end{equation}
Similarly, one can determine the intertwiner parameters for $j=1$
and analytically derive $SU(2)$-invariant  states with the maximal
entanglement entropy.

\end{document}